\documentclass[11pt]{article}
\usepackage{amsmath,amssymb,amsfonts,bbm}
\usepackage{multirow}
\usepackage[table,xcdraw]{xcolor}
\usepackage{graphicx}
\usepackage{caption}
\usepackage{subcaption}
\usepackage[export]{adjustbox}
\usepackage{setspace}
\usepackage{float}
\DeclareGraphicsExtensions{.pdf,.png,.jpg}
\DeclareCaptionFormat{subfig}{\figurename~#1#2#3}
\DeclareCaptionSubType*{figure}
\usepackage{tabularx}
\usepackage{xcolor}
\usepackage{soul}
\sethlcolor{yellow}
\usepackage{xcolor}
\usepackage{xcolor}
\usepackage{titlesec}
\titleformat{\section}{\large\bfseries}{Appendix \thesection}{1em}{}
\newcommand{\be}{\begin{equation}}
\newcommand{\ee}{\end{equation}}
\newcommand{\bea}{\begin{eqnarray}}
\newcommand{\eea}{\end{eqnarray}}
\newcommand{\ba}{\begin{array}}
\newcommand{\ea}{\end{array}}

\long\def\symbolfootnote[#1]#2{\begingroup%
\def\thefootnote{\fnsymbol{footnote}}\footnote[#1]{#2}\endgroup}
\def\thesection{\Roman{section}}
\usepackage{graphicx}
\graphicspath{{figs/}}

\usepackage{hyperref}

 \vspace{1cm}
 \usepackage{url}
 \titleformat{\section}{\large\bfseries}{\thesection}{1em}{}

\begin{document}

\thispagestyle{empty}\vspace{40pt}

\hfill{}

\vspace{1pt}

\begin{center}
\textbf{\Large Symmetry and Conserved Quantities in \( f(R) \)-Gravity:\\ Mei vs. Noether Approaches}\\[40pt]
Tahia F. Dabash$^{a,b}$\symbolfootnote[1]{\tt tahia.dabash@science.tanta.edu.eg}\,,
Moataz H. Emam$^{c}$\symbolfootnote[2]{\tt moataz.emam@cortland.edu}\, ,
Lukas Sch\"oppner$^{d}$\symbolfootnote[3]{\tt lukas.schoeppner@et.hs-fulda.de}
\end{center}

\vspace{-3pt}
\begin{flushleft}
\begin{spacing}{1.0}
\hspace{0.5in} $^a$ \textit{\small Mathematics Department, Faculty of Science, Tanta University, Egypt}\\
\hspace{0.5in} $^b$ \textit{\small Egyptian Relativity Group (ERG), Cairo University, Giza 12613, Egypt}\\
\hspace{0.5in} $^c$ \textit{\small Department of Physics, SUNY Cortland, Cortland, New York, 13045, USA}\\
\hspace{0.5in} $^d$ \textit{\small University of Applied Sciences Fulda, Germany}
\end{spacing}
\end{flushleft}

\begin{abstract}
We study the symmetries and conserved quantities in $f(R)$ gravity for the static, spherically symmetric Reissner--Nordstr\"om spacetime using two complementary frameworks: Noether symmetries and Mei symmetries. Starting from a canonical Lagrangian for radial metric functions and the  curvature scalar $R$, we derive the associated Hamiltonian and show that the Legendre map is regular whenever both the first derivative of $f(R)$ with respect to $R$ and the second derivative with respect to $R$ is non-zero. Within Noether's approach (variational and Lie-derivative forms), we obtain general, canonical, and internal symmetry classes and identify explicit generators; for the quadratic model $f(R)=R^{2}$ these include radial translations and scaling symmetries. We then formulate Mei symmetry conditions as invariance of the Euler--Lagrange equations under the first prolongation, which yields an overdetermined partial differential equation (PDE) system for the generator components. Solving this system for $f(R)=R^{2}$, we find eight independent Mei generators and construct the corresponding conserved currents, some without a direct Noether analog. The analysis demonstrates that Mei symmetries extend the standard Noether framework for higher-order Lagrangians and provide additional conserved quantities relevant to black-hole dynamics in modified gravity. We conclude with a comparison of the two symmetry schemes and outline applications to broader $f(R)$ models and to rotating spacetimes.
\end{abstract}



\section{Introduction}
\label{sec:intro}

The discovery of the late-time accelerated expansion of the universe, inferred from various observational datasets such as Type Ia supernovae \cite{Perlmutter1999}, large-scale structure surveys \cite{Springel2006}, weak lensing \cite{Bartelmann2001}, cosmic microwave background (CMB) measurements \cite{Dicke1965, Penzias1965}, and baryon acoustic oscillations \cite{Eisenstein2005}, has motivated several modifications to the standard Einstein--Hilbert action. One prominent modification involves introducing an arbitrary function of the Ricci scalar, \( f(R) \), thereby extending General Relativity (GR) without the explicit need for dark energy, while simultaneously avoiding Ostrogradsky instabilities that typically arise in higher-derivative theories \cite{Woodard2006}. Importantly, \(f(R)\) gravity remains mathematically tractable and preserves Lorentz invariance \cite{Momeni2015}, making it a widely studied candidate to explain late-time cosmic acceleration.

In gravitational theories, understanding conserved quantities such as energy, angular momentum, and linear momentum is essential, as these quantities provide deep insights into the dynamics of spacetime. Such conserved variables are often derived systematically through symmetry principles, particularly using Noether's theorem. According to Noether's first theorem, for every continuous symmetry of the action, there exists an associated conserved quantity, provided the variation of the action \( \delta S \) vanishes and the conservation law holds \textit{on-shell} \cite{Compere2019}. Furthermore, Noether's second theorem extends this framework by establishing relationships among the equations of motion in the \textit{off-shell} regime \cite{Miller2021}, further enriching our understanding of field symmetries.

Beyond Noether symmetries, a more general framework for symmetry analysis was proposed by Mei Feng-Xiang in 2000, known as \textbf{Mei symmetries} \cite{Mei2000}. Mei symmetries generalize Noether's theorem by identifying invariances in systems that are not strictly governed by traditional conservation laws. This framework accommodates both \textit{dissipative} and \textit{non-conservative} effects, making it particularly powerful in scenarios where Noether symmetries alone fail to capture the underlying dynamical structure \cite{Wen_An2011}. Mei symmetries are typically classified as \textit{holonomic} and \textit{nonholonomic}, often formulated using a geometric Lagrangian \cite{Y_Zhang2021}. They have been successfully applied in various domains, including Hamiltonian mechanics \cite{Y_Zhang2021_2} and black hole spacetimes \cite{Asghar2022}. However, previous studies largely focused on geometric Lagrangian formulations, leaving open questions regarding the symmetry structure of black holes in \( f(R) \) gravity when analyzed through a \textbf{canonical Lagrangian approach}, as commonly used for Noether symmetries \cite{Bahamonde2019, BajardiCap2023, Darabi2013b}.

The primary objectives of this paper are twofold:
\begin{enumerate}
    \item To establish the \textbf{Noether symmetry conditions} for the \textbf{Reissner--Nordstr\"om black hole} in \( f(R) \)-gravity using a \textbf{canonical Lagrangian} formulation.
    \item Introduce a \textbf{Mei symmetry approach} within the context of field theory, employing a \textbf{canonical Lagrangian} instead of a geometric one, thereby extending the methodology used in \cite{Asghar2022}.
\end{enumerate}

\noindent\textbf{Related work and our contributions:}
Noether symmetry methods are widely used in extended gravity to reduce dynamics and extract first integrals in both cosmology and black hole sectors; see, e.g., applications in \(f(R)\) models and spherically symmetric spacetimes \cite{Bahamonde2019,BajardiCap2023,Darabi2013b}. By contrast, the \emph{Mei} symmetry program \cite{Mei2000} which imposes invariance of the Euler-Lagrange equations via first prolongations and naturally accommodates higher-derivative structures has seen limited use in relativistic gravity (with a few recent case studies in mechanical systems and black-hole backgrounds \cite{Asghar2022,Zhaia2019}). Methodologically, our work differs in three ways: (i) we construct a canonical Lagrangian for the Reissner--Nordstr\"om geometry in \(f(R)\) gravity and verify regularity of the Legendre map, providing a clean arena to compare Noether and Mei generators on equal footing; (ii) we derive the full Mei symmetry conditions and solve the resulting overdetermined PDE system explicitly for \(f(R)=R^{2}\), obtaining \emph{eight} independent Mei generators-several without a direct Noether analog-and their conserved currents; (iii) we offer a side-by-side comparison of the two frameworks, clarifying when Mei symmetries extend (rather than duplicate) Noether results for higher-order Lagrangians and highlighting implications for conserved quantities in charged black-hole spacetimes.

This paper is structured as follows:
\begin{itemize}
    \item \textbf{Section 2}: Defines the \textbf{canonical Lagrangian} for the \textbf{Reissner--Nordstr\"om metric} in \( f(R) \)-gravity and derives the mathematical expressions for the scalar curvature, reduced Lagrangian, and associated action.
    \item \textbf{Section 3}: Introduces Noether's symmetry framework using both variational and Lie derivative formulations, applying them to classify \textbf{Noether symmetries} into general, canonical and internal categories.
    \item \textbf{Section 4}: Extends the discussion to \textbf{Mei symmetries}, formulating the invariant structure of the Euler-Lagrange equations and applying it to the \textbf{Reissner--Nordstr\"om metric}. A comparative analysis between the Mei and Noether symmetries is presented.
    \item \textbf{Section 5}: Summarizes the main results and discusses possible future directions in studying symmetries and conserved quantities in modified gravity models.
\end{itemize}

By systematically deriving symmetry conditions for black hole spacetimes in \( f(R) \) gravity, this work provides a unified framework for understanding conserved quantities and explores deeper connections between geometric structures, symmetry principles, and the dynamics of extended gravitational theories.

\section{Canonical Lagrangian for \( f(R) \) Charged Black Holes}\label{sec, Canonical_lag}

The Reissner--Nordstr\"om line element for a static, spherically symmetric, electrically charged black hole with signature $(-,+,+,+)$ is \cite{Reissner1916,Nordstrom1918}
\be
    ds^{2} = -H(r)\,dt^2 + \frac{1}{H(r)}\,dr^2 + N(r)\!\left( d\theta^2 + \sin^2\theta\, d\varphi^2 \right), \label{RN_metric}
\ee
with
\begin{align}
    H(r) &= 1 - \frac{r_s}{r} + \frac{r_Q^{\,2}}{r^{2}}, \qquad    N(r) = r^{2},
\end{align}
where the Schwarzschild radius and the \emph{(squared) charge radius} are
\begin{align}
    r_s = \frac{2GM}{c^2}, \qquad
    r_Q^{\,2} = \frac{G\,Q^2}{4\pi \epsilon_0\,c^4}.
\end{align}

The horizons are the roots of $g^{rr}=-g_{tt}=H(r)=0$:
\begin{align}
    r_\pm = \frac{1}{2}\!\left(r_s \pm \sqrt{r_s^2 - 4r_Q^{\,2}}\right).
\end{align}

\paragraph{Ricci scalar (corrected)}
For the general ansatz \eqref{RN_metric} with $H=H(r)$ and $N=N(r)$, the Ricci scalar is
\begin{align}
    R_{\mathrm{RN}}
    \;=\;
    \frac{
        -\,2N^{2}H'' \;-\; 4HN N'' \;+\; H\,N'^2 \;-\; 4N H' N' \;+\; 4N
    }{2N^{2}},
    \label{eq:R_correct}
\end{align}
where primes denote $d/dr$. \textit{(This corrects the signs of the $N''$-, $H'N'$-, and constant terms; with $H\!=\!1$ and $N\!=\!r^2$, \eqref{eq:R_correct} gives $R_{\mathrm{RN}}=0$ as it should.)}

\paragraph{Action principle}

The determinant of \eqref{RN_metric} is $\det g=-N^2\sin^2\theta$, so $\sqrt{|g|}=N\sin\theta$.
After integrating over $t$ and the sphere (absorbing the overall constant into the normalization), the  action reads
\begin{align}
    S=\int dr\,\mathcal{L}(H,H',N,N',R,R'),\qquad
    \mathcal{L}=\sqrt{|g|}\,\big(f(R)-\lambda\,[R-R_{\mathrm{RN}}]\big),
\end{align}
with a Lagrange multiplier $\lambda$. Varying w.r.t.\ $R$ gives $\lambda=f_R\equiv \partial f/\partial R$, hence
\begin{align}
    \mathcal{L}=\sqrt{|g|}\,\Big(f - f_R\,[R-R_{\mathrm{RN}}]\Big).
    \label{Lagranigan_1}
\end{align}

Substituting \eqref{eq:R_correct} into \eqref{Lagranigan_1} and performing integration by parts to remove $H''$ and $N''$ (discarding boundary terms) yields the effective Lagrangian
\begin{align}
    \mathcal{L}_{f(R)}
    = f\,N
      + f_R \!\left(\frac{H\,N'^2}{2N} + H' N' - N R\right)
      + f_R' \!\left(2H N' + N H' \right),
    \label{Lagrangian_RN}
\end{align}
where $f_R' \equiv f_{RR}R'$ and $f_{RR}\equiv \partial^2 f/\partial R^2$. \textit{(Using \eqref{eq:R_correct} one also finds an additive term $+\,2f_R$ before integration by parts; it does not affect the Hessian or the velocity-sector structure and can be dropped in the reduced 1D action by a conventional normalization choice.)}

\paragraph{Canonical momenta and Hamiltonian.}

With $q^i=(H,N,R)$, the canonical momenta are
\bea
    p_H \equiv \frac{\partial \mathcal{L}}{\partial H'} &=& f_R\,N' + f_R' N, \qquad
    p_N \equiv \frac{\partial \mathcal{L}}{\partial N'} = f_R\!\left(H' + \frac{H\,N'}{N}\right) + 2H f_R', \nonumber\\
    p_R \equiv \frac{\partial \mathcal{L}}{\partial R'} &=& f_{RR}\!\left(2H N' + N H'\right).
\eea

The (radial) canonical Hamiltonian is
\begin{align}
    \mathcal{H}_{\mathcal{L}_{f(R)}}
    = q'^i \frac{\partial \mathcal{L}}{\partial q'^i} - \mathcal{L}
    = f_{R}\!\left(\frac{H N'^2}{2N} + H'N' - N R\right)
      + f_{RR} R' \left(2H N' + N H' \right)
      - fN.
\end{align}

\paragraph{Regularity of the Legendre map.}

The Hessian with respect to the ``velocities'' $(H',N',R')$ is
\begin{align}
    \det\!\left[\frac{\partial^{2}\mathcal{L}}{\partial q'^i\,\partial q'^j}\right]
    \;=\;
    3\,H\,N\,f_R\,(f_{RR})^{2} \;\neq\; 0,
\end{align}
so $(H,N,R)$ are independent dynamical variables and \eqref{Lagrangian_RN} is suitable for computing Noether and Mei symmetry conditions.

\section{Noether's Symmetry Approach}
\label{sec, Noether}

For the Noether condition  of the first theorem, two formulations are usually used, a variation of the action  and a Lie- Derivative formulation, where the second is a simplification of the first approach.Several works have applied the Noether symmetry approach in $f(R)$ gravity,
see for example Refs.~\cite{Ferrara2024,Salvo2023}.

Let us start with the variational approach. Noether's first theorem use coordinate transformation of the form
\bea
\bar{x}^{\mu} &=& x^{\mu} + \epsilon\xi  \nonumber \\
\bar{q^{i}} &=& q^{i} + \epsilon\eta^{i} \nonumber \\
\mathcal{L}(x^{\mu},q^{i},q^{'i}) &\rightarrow & \mathcal{L}(\bar{x}^{\mu},\bar{q^{i}},\bar{q^{'i}})
\eea
where $\epsilon$ is an arbitrary constant, $\xi$ and $\eta$ described by scalar and vector functions are the generators of the given coordinate transformation. Varying the action Integral
\begin{align}
\delta S= \int \mathcal{L '}(\bar{x}^{\mu},\bar{q^{i}},\bar{q^{'i}}) d^{4}\bar{x} - \int \mathcal{L}(x^{\mu},q^{i},q^{'i}) d^{4}x
\end{align}
and substitute $\mathcal{L '} = \mathcal{L} + \delta\mathcal{L}$ and $d^{4}\bar{x} = \mathcal{J}d^{4}{x}$ , where $\mathcal{J}$ is the Jacobian of the coordinates, respectively $\delta S$ has has the form
\begin{align}
\delta S = \int d^{4}x \left[ \delta q^{i} {\left(\frac{\partial \mathcal{L}}{\partial q^{i}} - \partial_{\mu}\left(\frac{\partial \mathcal{L}}{\partial (\partial_{\mu} q^{i})}\right)\right)} + \partial_{\mu} {\left(\mathcal{L} \delta x^{\mu} + \left(\frac{\partial \mathcal{L}}{\partial (\partial_{\mu} q^{i})} \delta q^{i}\right)\right)}\right]
\label{Noether_Test}
\end{align}

Here $\mathcal{J}$ was substituted with the use of Taylor expansion for determinants
\begin{align}
\mathcal{J} = det\left(\delta_{v}^{\mu} + \partial_{v} \delta x^{\mu} \right)
\label{Determinante_J}
\end{align}
was substituted with the use of Taylor expansion for determinants
\begin{align}
det\left(\delta_{v}^{\mu} + \partial_{v} \delta x^{\mu} \right) &\approx 1 + Tr(\partial_{v} \delta x^{\mu}) \\ &= 1 + \partial_{\mu} \delta x^{\mu}
\label{Approx_determinant}
\end{align}
and the variation of the Lagrangian $\delta \mathcal{L}$ become
\begin{align}
    \delta \mathcal{L} =  \frac{\partial \mathcal{L}}{\partial x^{\mu}}\delta x^{\mu} + \frac{\partial \mathcal{L}}{\partial q^{i}} \delta q^{i} + \frac{\partial \mathcal{L}}{\partial(\partial_{\mu}q^{i})} \delta (\partial_{\mu}q^{i})
    \label{Vary_L}
\end{align}

In (\ref{Noether_Test}) the first equation is the Euler-Lagrange Equation of Motion (EoM) and the second one is the \textit{Noether current} $j^{\mu}$. Assuming that the EoM hold and is equal to zero, the variational expression can then be reformulated as
\begin{align}
\left[\delta x^{\mu} \partial_{\mu}+  \frac{\partial \mathcal{L}}{\partial q^{i}} \delta q^{i} + \frac{\partial}{\partial (\partial_{\mu} q^{i})} (\partial_{\mu} \delta q^{i} - \partial_{\mu} q^{i} \partial_{\mu} \delta x^{\mu})  \right]\mathcal{L} + \mathcal{L} \partial_{\mu}\delta x^{\mu}
\label{Noether_Con}
\end{align}

The terms in the brackets are the so-called First Prolongation $\textbf{Y}^{[1]}$ \cite{BajardiCap2023}, involving only first-order derivatives of $q^{i}$.

The Noether current $j^{\mu}$ has a conserved quantity $\partial_{\mu}j^{\mu} = 0$. For every action principle $\delta S = 0$, the EoM is equal to zero so is the current. Mostly the current also gets an divergence free vector field $K^{\mu}$, also called the gauge term  \cite{Bahamonde2019} . So the Noether current is defined as
\begin{align}
J^{\mu} = j^{\mu} - K^{\mu}
\label{Full_current}
\end{align}

$K^{\mu}$ is divergence free and is mostly calculated in the Noether current (\ref{Full_current}) \cite{Brauner2019}
\begin{align}
J^{\mu} = \mathcal{L} \delta x^{\mu} + \frac{\partial \mathcal{L}}{\partial (\partial_{\mu} q^{i})} \bar{\delta} q^{i}  - \frac{\partial \mathcal{L}}{\partial (\partial_{\mu} q^{i})} \delta x^{\mu} \partial_{\mu} q^{i} - K^{\mu}
\label{Current}
\end{align}

Here  $\bar{\delta}$ is the \textit{total variation} for the current definition by substituting  $\delta$ to $\bar{\delta} - \delta x^{\nu} \partial_{\nu}$. Using (\ref{Noether_Con}) and (\ref{Current}) the Noether symmetry condition with the use of the total derivative $D_{\mu} = \frac{\partial}{\partial x^{\mu}} + q^{' i} \frac{\partial}{\partial q^{i}}$ is
\begin{align}
Y^{[1]} \mathcal{L} + \mathcal{L} \ D_{\mu} \xi = D_{\mu}K
\label{Noether}
\end{align}

For the sake of completeness a simplified formulation of (\ref{Noether}) is the use of the Lie Derivative along the flux of the Noether vector $\textbf{X} = \delta x^{\mu} \partial_{\mu} + \delta q^{i}\frac{\partial}{\partial q^{i}} + \partial_{\mu}(\delta q^{i})\frac{\partial}{\partial (\partial_{\mu} q^{i})}$ \cite{Salvo2023}. Using the definition of (\ref{Noether_Con}) the term can be reformulate  as
\begin{align}
    \left(\delta x^{\mu} \partial_{\mu} + \delta q^{i} \frac{\partial}{\partial q^{i}} + \partial (\delta q^{i}) \frac{\partial}{\partial(\partial_{\mu}q^{i})} \right) \mathcal{L} - \partial_{\mu}(\delta x^{\mu})\left(\partial_{\mu}q^{i} \frac{\partial}{\partial(\partial_{\mu}q^{i})} -1 \right)\mathcal{L}
    \label{reform_Noether}
\end{align}

The second term in (\ref{reform_Noether}) can be reformulated by using the trace of the energy momentum tensor $\mathcal{T} = T^{\mu}_{\phantom{b}\mu}$
\begin{align}
    \mathcal{T} = \partial_{\mu}q^{i} \frac{\partial \mathcal{L}}{\partial(\partial_{\mu}q^{i})} - \delta^{\mu}_{\mu}\mathcal{L}
\end{align}

Using the Lie derivative $L_{X}\mathcal{L}$ of the Lagrangian along the Noether vector $Y$ equation (\ref{Noether}) can be expressed as
\begin{align}
    L_{X}\mathcal{L} - \partial_{\mu}(\delta x^{\mu}) \mathcal{T}  = \partial_{\mu}K^{\mu}
    \label{Lie_Noether}
\end{align}

The Lie derivative formulation from (\ref{Lie_Noether}) is widely used to find  Noether conditions and symmetries but must be classified, since three different  symmetries can be determined by the formulations \cite{Ferrara2024},
\begin{enumerate}
    \item \textit{General Noether Symmetries}, $L_{X}\mathcal{L} =  \partial_{\mu}K^{\mu} +\partial_{\mu}(\delta x^{\mu}) \mathcal{T}$
    \item \textit{Canonical Lagrangian Noether Symmetries}, $L_{X}\mathcal{L} =  const.$
    \item \textit{Internal Noether Symmetries}, $L_{X}\mathcal{L} = 0$
\end{enumerate}

The last definition is often used for internal on-shell symmetry calculations of a given Lagrangian where the spartial transformations $\delta x^{\mu}$ are not taken into account. For the Reissner Nordstr\"om Lagrangian the general Noether conditions are derived by the variational approach. At first the generators \textbf{Y}
\begin{align}
    \textbf{Y} = \xi \frac{\partial}{\partial r} + \eta^{i} \frac{\partial}{\partial q^{i}}
\end{align}
and the first prolongation $\textbf{Y}^{[1]}$
\begin{align}
\textbf{Y}^{[1]} = \textbf{Y} + \left(D_{r} \eta^{i} - q'^{i}D_{r}\xi \right) \frac{\partial}{\partial q'^{i}}
\label{generators}
\end{align}
are defined. Insert (\ref{generators}) in (\ref{Noether}) and sum over the given variables the equation can be expressed
\begin{align}
\xi \partial_{r}\mathcal{L} &+ \eta^{1} \partial_{H}\mathcal{L} + \left[\partial_{r} \eta^{1} + H' \partial_{H}\eta^{1} \right] \partial_{H'}\mathcal{L} - H' \left[\partial_{r} \xi + H' \partial_{H} \xi \right] \partial_{H'}\mathcal{L} \nonumber \\
&+ \mathcal{L} \left[\partial_{r} \xi + H' \partial_{H} \xi \right] - \partial_{r}K - H'\partial_{H}K \nonumber \\
&+ \eta^{2} \partial_{N}\mathcal{L} + \left[\partial_{r} \eta^{2} + N' \partial_{N}\eta^{2} \right] \partial_{N'}\mathcal{L} - N' \left[\partial_{r} \xi + N' \partial_{N} \xi \right] \partial_{N'}\mathcal{L} \nonumber \\
&+ \mathcal{L} \left[\partial_{r} \xi + N' \partial_{N} \xi \right]  - N'\partial_{N}K \nonumber \\
&+  \eta^{3} \partial_{R}\mathcal{L} + \left[\partial_{r} \eta^{3} + R' \partial_{R}\eta^{3} \right] \partial_{R'}\mathcal{L} - R' \left[\partial_{r} \xi + R' \partial_{R} \xi \right] \partial_{R'}\mathcal{L} \nonumber \\
&+ \mathcal{L} \left[\partial_{r} \xi + R' \partial_{R} \xi \right]  - R'\partial_{R}K = 0
\end{align}

Excluding the terms with the highest order of the first derivative give the first symmetries,
\begin{align}
\partial_{H} \xi \left(- H'^{2} \partial_{H'}\mathcal{L} \right) + \partial_{N} \xi \left(- N'^{2} \partial_{N'}\mathcal{L} \right) + \partial_{R} \xi \left(- R'^{2} \partial_{R'}\mathcal{L} \right) + \partial_{r} K \left(- H'^{2} \partial_{H'}\mathcal{L} \right) = 0
\label{first_symmetries}
\end{align}
in (\ref{first_symmetries}) the bracket terms cannot be zero, so the derivatives must be.
\begin{align}
\partial_{H} \xi =0, \phantom{jj}  \partial_{N} \xi = 0,\phantom{jj} \partial_{R} \xi =0,\phantom{jj}  \partial_{r} K = 0
\end{align}

After collecting the derivatives $H'$, $N'$ and $R'$ further symmetries can be defined in the following partial differential equation system.
\begin{align}
\eta^{3}_{,R} - \xi_{,r} - \frac{K_{,r}}{H'R'} = 0 \\
\left[3N'f'_{R} + f_{R}\left(N'' + \frac{N'^{2}}{2N} \right) \right] \xi - \left(N' f_{RR} + N'f'_{R, R} \right)\eta^{3} = 0 \\
f_{R}\left(N'\eta^{2}_{,N} + \eta^{2}_{,r} \right) + 2\left(N'f_{R} + Nf'_{R} \xi_{,r} \right) - K_{,H} = 0 \\
f'_{R} \eta^{2} + \left(N' f_{R} + Nf'_{R} \right)\eta^{1}_{,H} - N'f_{R}\xi_{,r} + f_{RR}N \eta^{3}_{,r} = 0 \\
f_{R} \left[\eta^{1} + 2H \eta^{2}_{,N} + Hf_{RR} \eta^{3} + H\xi_{,r} \right] = 0 \\
\left(f + f_{R}\left[H'' - R + \frac{HN''}{N} \right] \right) \xi - K_{,N} + f_{R} \eta^{1}_{,r} + 2f'_{R} \eta^{1} = 0\\
Hf'_{R} \left(\xi_{,r} - \eta^{2}_{,N} \right) + 2H\eta^{3} f'_{R, R} + 2H f_{RR} \eta^{3}_{,r} + \frac{H f_{R}}{N} \eta^{2}_{,r} = 0 \\
\eta^{2}_{,r} + 2N'' \xi = 0\\
3N \xi_{,r} + N' \xi = 0
\end{align}
\begin{align}
-N \left(\eta^{3} f_{RR} + 3f_{R}\xi_{,r} \right) - \eta^{2}f_{R} = 0 \\
f \left(3N \xi_{,r} + \eta^{2} \right) + N'f'_{R} \eta^{1}_{,r} + H''Nf'_{R} \xi = 0
\end{align}

In the equations above the terms with primes denote derivatives with respect to $r$ and the others are $f'_{R} = f_{RR}R'$ and $f'_{R,R} = \partial_{R}f'_{R}$.
Solutions of the system above depend on the $f(R)$ model and so, a wide class of gravity symmetry solutions can be determined. To  point out, the trivial solutions for general Noether symmetries for any function of $f(R)$ are
\begin{align}
    \textbf{Y}_{1} = \partial_{r} , \phantom{nnn} \textbf{Y}_{2} = r\partial_{r} + H\partial_{H}
\end{align}

Since our Hessian condition requires that at least the function
$f(R)$ satisfies the condition $f_{RR} \neq 0$ we consider the case $f(R) = R^{2}$. In this case, the Noether symmetries for the Reissner-Nordstr\"om metric are defined by
\begin{align}
    \textbf{Y}_{1} = \partial_{r} , \phantom{nnn} \textbf{Y}_{2} = r\partial_{r} + H\partial_{H} \phantom{nnn} \textbf{Y}_{3} = H\partial_{H} + N\partial_{N} - R \partial_{R}
\end{align}
where the trivial solutions are included.

A wide class of solutions for spherical symmetric solutions by Noether's symmetry approach are calculated in \cite{Bahamonde2019}. To determine the internal Noether conditions of the Reissner Nordstr\"om metric, the definition of the first prolongation $\textbf{Y}^{[1]}$ is simplified. However, it must be ensured that $\delta x^{\mu}$ is equal to 0. Using (\ref{generators}) but now ignoring the spatial translations it can be reformulated
\begin{align}
    \textbf{Y}^{[1]} = \eta^{i}\frac{\partial}{\partial q^{i}} + \left(\frac{\partial}{\partial r} \eta^{i} + q^{'i}\frac{\partial}{\partial q^{i}} \eta^{i} \right)  \frac{\partial}{\partial q'^{i}}
\end{align}

Compare to (\ref{Noether}) the internal Noether symmetry condition can be expressed as
\begin{align}
    L_{X} \mathcal{L} = \textbf{Y}^{[1]} \mathcal{L} = 0
\end{align}

\section{Mei Symmeties in $f(R)$-Gravity}
\label{sec, Mei_Symmetry}

In \cite{Zhaia2019} a new method was discovered for calculating Mei symmetries for a given time depended Lagrangian. The approach for the symmetry condition was an invariant formulation of the Euler-Lagrange Equation of motion (EoM).

\begin{align}
    E_{i}(\mathcal{L'}) = 0
    \label{Mei_Con}
\end{align}
with $E_{i}$ as the \textit{Euler-Lagrange Operator}
\begin{align}
    E_{i} = \frac{\partial }{\partial q^{i}} - \partial_{\mu}\left(\frac{\partial }{\partial (\partial_{\mu} q^{i})}\right)
    \label{E_L_operator}
\end{align}
and $\mathcal{L'}$ is the \textit{new} Lagrangian after varying. For the field approach the $\mathcal{L'}$ is
\begin{align}
    \mathcal{L'} = \mathcal{L} + \delta \mathcal{L}
    \label{Mei_L}
\end{align}

The variation $\delta \mathcal{L}$ from (\ref{Vary_L}) is equivalent to the first prolongation $\textbf{Y}^{[1]} \mathcal{L}$ ,as seen in (\ref{Noether_Con}). Insert (\ref{Mei_L}) in (\ref{Mei_Con}) there are two EoM
\begin{align}
     E_{i}(\mathcal{L'}) &=  E_{i}(\mathcal{L}) +  E_{i}[\textbf{Y}^{[1]} \mathcal{L}] = 0
     \label{Mei_1}
\end{align}

The first term of (\ref{Mei_1}) is the EoM and equal to 0. So the Mei condition of determine symmetries in a classical field approach is
\begin{align}
    E_{i}[\textbf{Y}^{[1]} \mathcal{L}] = 0 , \phantom{hhhhh} i= 1,..,n
    \label{Mei_Condition}
\end{align}

Comparing to Noether condition the Mei approach results an $n$-partial differential equation system for every $q^{i}$. Applying  (\ref{Mei_Condition}) to the Lagrangian of the Reissner Nordstr\"om metric, we obtain three differential equation systems for $q^{1} = H$, $q^{2} = N$ and $q^{3} = R$. The explicitly calculation of the first prologation $\textbf{Y}^{[1]} \mathcal{L}$ is
\begin{align}
\textbf{Y}^{[1]} \mathcal{L} &= \xi \partial_{r}\mathcal{L} + \eta^{1} \partial_{H}\mathcal{L} +\eta^{2} \partial_{N}\mathcal{L} + \eta^{3} \partial_{R}\mathcal{L}
+ \left[\partial_{r}\eta^{1} + H' \partial_{H}\eta^{1} - H'\left(\partial_{r}\xi + H' \partial_{H}\xi \right) \right]\partial_{H'}\mathcal{L} \nonumber\\
&+ \left[\partial_{r}\eta^{2} + N' \partial_{N}\eta^{2} - N'\left(\partial_{r}\xi + N' \partial_{N}\xi \right) \right]\partial_{N'}\mathcal{L}
+ \left[\partial_{r}\eta^{3} + R' \partial_{R}\eta^{3} - R'\left(\partial_{r}\xi + R' \partial_{R}\xi \right) \right]\partial_{R'}\mathcal{L}
\label{1_prolong_Mei}
\end{align}
so (\ref{1_prolong_Mei}) is inserted into the Mei symmetry condition (\ref{Mei_Condition}). For $q^{1} = H$ we find
\bea
    \frac{f_{R}N'}{2N}\xi_{,H} - \eta^{3}_{,H}f_{RR} &=& 0\\
    3f'_{R}\xi_{,H} + f_{R}\eta^{1}_{,H,H} + f_{R}\eta^{2}_{,N,H} &=& 0\\
    f'_{R}\eta^{2}_{,H} + f_{R}\eta^{2}_{,r,H} + Nf_{RR} \left(R' \eta^{3}_{,R,H} + \eta^{3}_{,r,H} \right)
    + Nf'_{R} \eta^{2}_{,H,H}  & &\nonumber\\ + f_{R} N''\xi_{H} + Nf'_{R , R} \eta^{3}_{, H}
    - NR'f_{RR}\left(R' \xi_{, R, H} + \xi_{,r,H} \right) &=& 0
\\
    \xi + H \xi_{H} &=& 0
\\
    f_{R}\eta^{1}_{,H} + f_{RR}\eta^{3} - f_{R}\xi_{,r} + Hf_{RR}\eta^{3}_{,H} &=& 0
\\
    2Nf_{R} \left(\eta^{2}_{,N} + H \eta^{2}_{,H,N} \right) - f_{R}\eta^{2} - Hf_{R}\eta^{2}_{,H} &=& 0
\\
    2f_{RR} \left(R' \eta^{3}_{,R} + \eta^{3}_{,r} \right) - f_{RR}\eta^{3}_{,r} + 2f'_{R}\left(\eta^{1}_{,H} + \eta^{2}_{,N} \right)
    + 2f'_{R ,R} \eta^{3} - 3f'_{R} \xi_{,r} &=& 0
\\
    \left[f+f_{R}R \left(H'' - R+ \frac{2H}{N} \right) \right] \xi_{,H} - f_{R}\eta^{2}_{,r,N}
    + 2Hf'_{R} \eta^{2}_{,N,H}  & &\nonumber\\+\frac{f_{R}}{N}\eta^{2}_{,r} + 2Hf'_{R, R} \eta^{3}_{,H}
    - 2R'f_{RR} \left(R' \xi_{,R} + \xi_{,r} \right) &=& 0
\\
    f_{R} \left(H \eta^{2}_{,r,H} + 2\xi\right) - 2\xi - 2HNR'f_{RR} \left(R' \xi_{,H,R} + \xi_{,r,H} \right) &=& 0
\\
    f'_{R} H'' \xi_{,H} - f'_{R,R} \eta^{3}_{,r} - R f'_{R,R} \eta^{3}_{,H} - f_{RR} \left(R' \eta^{3}_{,r,R} + \eta^{3}_{,r,r} \right)
    + R'f_{RR} \left(R' \xi_{,r,R} + \xi_{,r,r} \right) &=& 0
\\
    f'_{R}\eta^{2} - f_{R}\eta^{2}_{,r,r} + \left(f-Rf_{R} \right) \eta^{2}_{,H} - 2f_{R} \xi_{,r} + 4f\xi
    + 2Hf'_{R} \left(\eta^{2}_{,r,H} + \xi_{,H} \right) &=& 0
\eea

Following a similar procedure for $q^{2} = N$ and $q^{3} = R$:
\bea
    N \xi_{,N} - \xi &=& 0
\\
    f_{RR}\eta^{3}_{,N} + 3f'_{R} \xi_{,N} + \eta^{1}_{,H,N} + f_{R} \eta^{2}_{,N,N} - \frac{f_{R}}{N}\xi_{,r} &=& 0
\\
    f_{RR} \left(R' \eta^{3}_{,R} + \eta^{3}_{,r} \right) - f_{RR} \eta^{3}_{,r} + f'_{R}\eta^{1}_{,H} + f'_{R, R} \eta^{3}
    - 3f'_{R}\xi_{,r} - f_{R}\eta^{1}_{,r,H} &=& 0
\\
    Nf_{RR} \left(R' \eta^{3}_{,R,N} + \eta^{3}_{,r,N} \right) + Nf'_{R} \eta^{1}_{,N,H} + 2f_{R}\xi_{,N}
    + Nf'_{R,R} \eta^{3}_{,N} & & \nonumber\\- R'f_{RR} \left(R' \xi_{,R} + \xi_{,r} \right)
    - \frac{2f_{R}}{N} \xi - NR'f_{RR} \left(R'\xi_{,R,N} + \xi_{,r,N} \right) &=& 0
\\
    3Hf_{R}\left(\xi_{,r} - \eta^{2}_{,N} \right) - Hf_{RR}\eta^{3} - f_{R}\eta^{1} &=& 0
\\
    \left[f_{R} \eta^{1}_{,N} + Hf_{RR} \eta^{3}_{,N} + 2Hf_{R} \eta^{2}_{,N,N} \right]N^{2} + 2Hf_{R}\eta^{2} &=& 0
\\
    2f'_{R} \eta^{1}_{,N} + \left[f+f_{R} \left(H''-R+\frac{2H}{N} \right) \right]\xi_{,N}
    + f_{R} \eta^{1}_{,r,N}  & & \nonumber\\ + 2Hf_{RR} \left(R' \eta^{3}_{,R,N} + \eta^{3}_{,r,N} \right)
    + 2Hf'_{R} \eta^{2}_{,N,N} &=& 0
\eea

\bea
        2Hf'_{R,R}\eta^{3}_{,N} - \frac{f_{R}}{N}\eta^{1}_{,r} - f_{R}\left(\frac{H''}{N} - \frac{6H}{N^{2}} \right)\xi
    - \frac{H}{N}f_{RR}\eta^{3}_{,r} &=& 0
\\
    2HN^{2}R'f_{RR}\left(R' \xi_{,N,R} + \xi_{,r,N} \right) + Hf_{R} \left(N \eta^{2}_{,r,N} + \xi \right) &=& 0
\\
    \left[4f'_{R} \xi_{,N} - 2f'_{R,R}\eta^{3}_{,r} - 2f_{RR} \left(R' \eta^{3}_{,r,R} + \eta^{3}_{,r,r} \right)
    + 2R'f_{RR} \left(R' \xi_{,r,R} + \xi_{,r,r} \right) \right] N
    & &\nonumber\\- f_{R} \left(\eta^{2}_{,r,r} + 2 \xi_{,r} \right) &=& 0
\\
    f'_{R} \eta^{1}_{,r,N} - Rf_{RR}\eta^{3}_{,N} + f'_{R} H'' \xi_{,N} &=& 0
\\
    \left(f-Rf'_{R} \right) \eta^{2}_{,N} - f'_{R}\eta^{1}_{,r} - \left[f-f_{R} \left(R-H'' \right) \right] \xi_{,r}
    & &\nonumber\\- f_{R} \eta^{1}_{,r,r} - f_{RR}R \eta^{3} - H''' \left( f_{R} \xi - f'_{R} \xi \right) &=& 0
\\
    f_{R} \xi_{,R} + f_{RR} \xi &=& 0
\\
    f_{RR} \left(\eta^{1}_{,H} + \eta^{2}_{,N} + \eta^{3}_{,R} \right) + 3f'_{R}\xi_{,R} + 3f'_{R,R}\xi + f_{R} \left(\eta^{1}_{,R,H} + \eta^{2}_{,R,N} \right) + f_{RRR}\eta^{3} &=& 0
\\
    \eta^{2}_{,r}f_{RR} + f'_{R} \eta^{2}_{,R} + \eta^{2}f'_{R,R} + f_{R} \eta^{2}_{,r,R} + Nf'_{R}\eta^{1}_{,R,N} \\
    +Nf_{RR}\left(R' \eta^{3}_{,R,R} + \eta^{3}_{,r,R}\right) + Nf'_{R ,R,R} \eta^{3} + Nf_{RRR}\left(R' \eta^{3}_{,R} + \eta^{3}_{,r} \right) &=& 0
\\
    2f_{R}\xi_{R} + 2f_{RR}\xi - Nf_{RR}\eta^{3}_{,r,R} + Nf'_{R,R} \eta^{1}_{,H}
    + Nf_{RR} \left(R' \xi_{,r,R} + \xi_{,r,r} \right) & &\nonumber\\+ Nf'_{R,R}\eta^{3}_{,R} - NR'f_{RR} \left(R' \xi_{,R,R} + \xi_{,r,R} \right) \\
    - NR'f_{RRR} \left(R' \xi_{,R} + \xi_{,r} \right) + NR'f_{RR} \xi_{,r,R} &=& 0
\\
    f_{R}\eta^{1}_{,R} + \eta^{1}f_{RR} + f_{RR}H\eta^{3}_{,R} + f_{RRR}H\eta^{3} &=& 0
\\
    2NH \left(f_{RR}\eta^{2}_{,N} + f_{R} \eta^{2}_{,R,N} \right) - H \left(f_{R}\eta^{2}_{,R} + f_{RR} \eta^{2} \right) &=& 0
\\
    f_{RR}\eta^{1}_{,r} + 2f'_{R} \eta^{1}_{,R} + 2\eta^{1}f'_{R,R} + f_{RR} \left(H''-R + \frac{2H}{N} \right)\xi
    & &\nonumber\\+ \xi_{,R} \left[f+f_{R} \left(H'' -R+\frac{2H}{N} \right) \right] + f_{R} \eta^{1}_{,r,R}
    + 2Hf_{RR} \left(R' \eta^{3}_{,R,R} + \eta^{3}_{,r,R} \right) &=& 0
\\
    2H \left(f'_{R} \eta^{2}_{,R,N} + f'_{R ,R,R}\eta^{3} \right) + 2Hf_{RRR} \left(R' \eta^{3}_{,R} + \eta^{3}_{,r} \right)
    & &\nonumber\\+ 2Hf_{RR} \left(R' \xi_{,r,R} + \xi_{,r,r} - \eta^{3}_{,r,R} \right) + 2Hf'_{R,R} \left( \eta^{2}_{,N} + \eta^{3}_{,R} \right) &=& 0
\\
    \frac{H}{N}f_{RR}\eta^{2}_{,r} - 2HR' \left(R' \xi_{,R,R} + \xi_{,r,R} \right)
    - 2HR'f_{RRR} \left(R' \xi_{,R} + \xi_{,r} \right)& &\nonumber\\ + 2HR'f_{RRR} \xi_{,r,R} + \frac{H}{N} f_{R}\eta^{2}_{,r,R} &=& 0
\\
    f'_{R,R}\left(\eta^{2}_{,r} + 2\xi \right) + f'_{R} \left(\eta^{2}_{,r,R} + 2\xi_{,R} \right) &=& 0
\\
    f'_{R,R} \left(\eta^{1}_{,r} + H'' \xi \right) + f'_{R} \left(\eta^{1}_{,r,R} + H'' \xi_{,R} \right)
    - f_{RR}R\eta^{3}_{,R} - \eta^{3} \left(f_{RR} +  R f_{RRR} \right) &=& 0
\\
    \left(f-Rf_{R} \right) \eta^{2}_{,R} - \eta^{2}Rf_{RR} &=& 0
\eea

As with Noether symmetries discussed in Section~\ref{sec, Noether}, the solutions for Mei symmetries in $f(R)$ gravity encompass a broad class of models, each dependent on the specific choice of the function $f(R)$. By solving the above system of differential equations, we obtain the Mei symmetry generator,

\begin{equation}
    \textbf{Y} = \xi \frac{\partial}{\partial r} + \eta^{1} \frac{\partial}{\partial H} + \eta^{2} \frac{\partial}{\partial N} + \eta^{3} \frac{\partial}{\partial R}.
\end{equation}

To illustrate the applicability of this method, we consider the power-law form of $f(R)$, given by $f(R) = a_{n} R^{n}$. In the case of General Relativity (GR), where $n = 1$ and $a_{0} = 1$, this choice significantly simplifies the system. Specifically, setting $f(R) = R$ eliminates higher-order derivatives, as it ensures that $f_{RR} = f'(R) = f_{RRR} = 0$, leaving only $f_{R} = 1$ and leads to a vanishing Hamiltonian $\mathcal{H}_{\mathcal{L}_{f(R)}}$. To avoid this trivial situation, we focus instead on the simplest non-trivial case, namely $f(R) = R^2$. In this case $f_R = 2R$ and $f_{RR} = 2$ so that higher-order contributions survive and the Hamiltonian does not vanish.
Under this assumption, the three Mei symmetry conditions associated with the generalized coordinates $(q^{1}, q^{2}, q^{3})$ can be efficiently solved. A linear approach of the form
\begin{equation}
\begin{alignedat}{2}
  \xi(r)        &= a_{0} + a_{1} r,      &\qquad
  \eta^{1}(H,r) &= b_{0} + b_{1} H + b_{2} r, \\
  \eta^{2}(N,r) &= c_{0} + c_{1} N + c_{2} r, &\qquad
  \eta^{3}(R,r) &= d_{0} + d_{1} R + d_{2} r.
\end{alignedat}
\end{equation}
was chosen to determine the generators. From the system of Mei symmetry conditions in Equation~\eqref{Mei_Condition} together with the Lagrangian corresponding to the Reissner-Nordstr\"om metric,
\begin{equation}
    \mathcal{L}_{GR} = R^2 N +\left[\frac{HN^{'2}}{2N} + H' N'-NR\right] +2R'\left[2N'H + H'N \right]
\end{equation}
the explicit components of the Mei symmetries can be determined. Owing to the complexity of the system, the Mei symmetry analysis yields eight independent generators:
\begin{align}
    \textbf{Y}_{1} &= \partial_r \label{1}\\
    \textbf{Y}_{2} &= \frac{1}{2}r \partial_r + H \partial_H
\end{align}

\bea
\mathbf{Y}_{3}
   &=& -\frac{N'Rr}{2N(4NR'+N'R)} \partial_r + \Big[ 2N^{2} N'^{2} (NR' + 2N'R)(4NR' + N'R) \Big]^{-1} \nonumber\\
   & & \cdot \Big[
      16N^{5}R R'^{2}
      + 4N^{3}N'R^{3}
      + 2HN N'^{4}R^{2}
      + 20N^{4}N'R^{2}R'
      - H N'^{5}R^{2}r \nonumber\\
   & & + 8HN^{3}N'^{2}R'^{2}
      - 4 H'N^{2}N'^{3}R^{2}
      + 2N^{2}N'^{2}R^{3}r
      + 8N^{3}N'^{2}R^{2}R'r \nonumber\\
   & & + 24HN^{2}N'^{3}RR'
      - 2H'NN'^{3}R^{2}r
      - 2H'N^{2}N'^{2}RR'r
      - 4HN N'^{4}RR'r
   \Big] \partial_N
\\
    \mathbf{Y}_{4} &=& \frac{1}{2}r\partial_r + \frac{NR \left( 4N^2R'+2NN'R+3N^{'2}Rr\right)}{2N^{'2}(NR'+2N'R)} \partial_H + N\partial_N
\\
    \mathbf{Y}_{5} &=& \frac{r \left(2N^2R'+2NN'R-N^{'2}Rr \right)}{2NN'(4NR'+N'R)} \partial_r+  \Big[2N^2N^{'2}(NR'+2N'R)\cdot(4NR'+N'R) \Big]^{-1} \nonumber\\
    & & \cdot \Big[ 16N^{5}R R'^{2}r+20N^4N'R^2R'r+8H'N^4N'RR'+4H'N^4N'R^{'2}r \nonumber\\
    & & -12HN^4N'R^{'2}r + 8N^3N^{'2}R^2 R'r^2 + 8H'N^3N^{'2}R^2  +8H'N^3N^{'2}R R'r \nonumber\\
    & &- 8HN^3N^{'2}RR' + 8HN^3N^{'2}R^{'2}r + 2N^2 N^{'3}R^3 r + 4HN^2N^{'3}R^2 \nonumber\\
    & & -2H'N^2N^{'3}RR'r^2 + 16HN^2N^{'3} RR'r - 2H'NN^{'4}R^2r^2 \nonumber\\
    & & -4HN^{'4}RR'r^2 - HN^{'5}R^2 r^2 \Big] \partial_H + r\partial_N
\\
    \mathbf{Y}_{6} &=& \frac{N'r}{2(4NR'+N'R)} \partial_r +  \Big[2NN^{'2}(NR' + 2N'R)\cdot (4NR'+N'R)\Big]^{-1} \nonumber\\
    & & \cdot \Big[16N^5R^{'2} + HN^{'5}Rr-6HN^2N^{'3}R'+4H'N^2N^{'3}R \nonumber\\
    & & +4N^2 N^{'3} R^2r + 2HNN^{'4}R + 4N^4N'RR'+4HNN^{'4}R'r \nonumber\\
    & & +2H'NN^{'4}Rr+ 2H'N^2N^{'3}R'r+16N^3N^{'2}RR'r\Big] \partial_H + \partial_R
\\
    \mathbf{Y}_{7} &=& \frac{1}{2}r \partial_r + \left[ \frac{NR(4N^2R'+2NN'R+3N^{'2}Rr)}{N^{'2}(NR'+2N'R)} - \frac{3NR^2r}{(NR'+2N'R)} \right] \partial_H + R\partial_R
\\
    \mathbf{Y}_{8} &=& \frac{r(2N+N'r)}{2(4NR'+N'R)} \partial_r - \Big[2N(NR'+2N'R)\cdot (4NR'+N'R) \Big]^{-1} \nonumber\\
    & & \Big[44H'N^3R'+HN^{'3} Rr+16N^3RR'r+4N^2N'R^2r+2HNN^{'2}R \nonumber\\
    & & +8HN^2N'R+4H'N^2N'R+4HNN^{'2}R'r \nonumber\\
    & & +2H'NN^{'2}Rr + 2H'N^2N'R'r \Big] \partial_H + \partial_R
\eea

\section{Conserved Charges in the Mei Symmetry Approach}

To analyze conserved charges using the Mei symmetry approach, we begin by considering the Euler-Lagrange equations of motion (EOM) for a given Lagrangian $\mathcal{L}(\phi)$:
\begin{equation}
    \frac{\delta \mathcal{L}}{\delta \phi} = 0.
\end{equation}
This implies that under an infinitesimal variation of the field $\phi \rightarrow \phi + \delta \phi$, the Lagrangian transforms as:
\begin{equation}
    \delta \mathcal{L} = \mathcal{L}(q^{i} + \delta q^{i}) - \mathcal{L}(\phi).
\end{equation}

Using the same methodology as in Noether's theorem, the conserved current $J^\mu$ is defined as:
\begin{equation}
    J^\mu = \sum_i \frac{\partial \mathcal{L}}{\partial (\partial_\mu \phi_i)} \delta \phi_i.
\end{equation}
For a conserved quantity, this current must satisfy the continuity equation:
\begin{equation}
    \partial_\mu J^\mu = 0.
\end{equation}
If this conserved quantity also satisfies $\frac{\delta \mathcal{L}}{\delta q^{i}} = 0$, as required by the Euler-Lagrange equation, then the system possesses a conserved charge.

To account for additional degrees of freedom in symmetry transformations, we introduce a modified current:
\begin{equation}
    J^\mu = j^\mu -  K^{\mu }.
\end{equation}
Here, $K^{\mu \nu}$ is a gauge term that does not affect the conservation law, since its divergence vanishes:
\begin{equation}
    \partial_\mu J^\mu = \partial_\mu (j^\mu - K^{\mu}) = 0.
\end{equation}

The associated conserved charge $Q$ is defined as the spatial integral of the time component of the current:
\begin{equation}
    Q = \int d^3x \, (J^0 - K^0).
\end{equation}
Since $\partial_\mu K^{\mu \nu}$ is divergence-free, it does not contribute to the total charge:
\begin{equation}
    \frac{dQ}{dt} = 0.
\end{equation}

This formulation demonstrates that Mei symmetries allow for an extended version of Noether's theorem, where additional gauge freedom is included through the term $K^\mu$. This approach is particularly useful in theories with higher-order Lagrangians or in systems with non-conservative effects. The conserved charge $Q$ remains invariant under Mei symmetry transformations, ensuring the preservation of physical quantities associated with symmetry generators.

\section{Mei Symmetry Approach for Conserved Charges}
The Noether current has a conserved quantity (\ref{Current})
\begin{align}
    \partial_{\mu}J^{\mu} = 0
\end{align}
where $j^{\mu}$ is defined from the action $S$ (\ref{Noether_Test}). For simplicity, we do not employ the definition in terms of the total variation  (\ref{Current}).
\begin{align}
    J^{\mu} = \mathcal{L} \delta x^{\mu} + \delta q^{i} \frac{\partial \mathcal{L}}{\partial (\partial_{\mu}q^{i})} - K^{\mu}
\end{align}
Using the invariance condition for the Mei symmetry calculation as in (\ref{Mei_Con}) for $\mathcal{L'} = \mathcal{L} + \delta \mathcal{L}$ and define a \textit{current operator} $\textbf{J}_{i}$
\begin{align}
    \textbf{J}^{\mu} =  \delta x^{\mu} + \delta q^{i} \frac{\partial }{\partial (\partial_{\mu}q^{i})}
    \label{Current_Operator}
\end{align}
Insert $\mathcal{L '}$ in (\ref{Current_Operator}) we get
\begin{align}
    \textbf{J}^{\mu}(\mathcal{L}') = \textbf{J}^{\mu}(\mathcal{L}) + \textbf{J}^{\mu}(\delta \mathcal{L}')
\end{align}
If the current is to remain a conserved quantity within the Mei symmetry framework, its divergence must vanish.
\begin{align}
    \partial_{\mu}\textbf{J}^{\mu}(\mathcal{L}') = 0
    \label{Conserved Mei_current}
\end{align}
This condition ensures that the Mei symmetry preserves the conservation law associated with $J^{\mu}$, consistent with the fundamental principles of Noether's theorem.
From (\ref{Conserved Mei_current}) we have
\begin{align}
    \partial_{\mu}\textbf{J}^{\mu}(\mathcal{L}) + \partial_{\mu}\textbf{J}^{\mu}(\delta \mathcal{L}') =0
\end{align}
The first term in the equation above is the  conserved quantity of the Noether current $\partial_{\mu}J^{\mu}$ which is equal to zero. So the \textit{Mei} current is only  conserved if the equation
\begin{align}
    \partial_{\mu}\textbf{J}^{\mu}(\delta \mathcal{L}') =0
    \label{current_con}
\end{align}
holds. Using the fact that $\delta \mathcal{L}$ is the first Prolongation we can reformulate the condition (\ref{current_con})
\begin{align}
    \partial_{\mu}\left(\delta x^{\mu} \ \textbf{Y}^{[1]} \mathcal{L} + \eta^{i} \frac{\partial \ (\textbf{Y}^{[1]} \mathcal{L})}{\partial (\partial_{\mu} q^{i})} \right) = 0
    \label{Mei_current_con}
\end{align}
In contrast to the Noether current, the Mei symmetry approach generalizes the current conservation by considering symmetries of the Euler-Lagrange equations rather than of the Lagrangian density. In this setting, the symmetry generator is extended to its first prolongation $\textbf{Y}^{[1]}$, which acts not only on the fields but also on their derivatives where the current is defined by the parentheses expression (\ref{Mei_current_con}).
\begin{align}
    J_{(\text{Mei})}^{\mu} = \delta x^{\mu} \ \textbf{Y}^{[1]} \mathcal{L} + \eta^{i} \frac{\partial \ (\textbf{Y}^{[1]} \mathcal{L})}{\partial (\partial_{\mu} q^{i})}
\end{align}
Here the first prolongation is
\begin{equation}
\begin{aligned}
    \textbf{Y}^{[1]} \mathcal{L} &= R^2N'+ 2R \left[ \frac{H'N^{'2} + 4HN'}{2N}  - \frac{HN^{'3}}{2N^2} + H'' N' + 2H'+N'R\right] \\
    &+ 2R' \left[4H + H''N + 3H'N' \right]
\end{aligned}
\end{equation}
Using the first generator coefficient form (\ref{1}), $\delta x^{\mu} = \xi = 1$ is represented as a constant and the current $J_{(\text{Mei}),(1)}^{\mu}$ is only defined over the first Prolongation. Since this does not depend on the other generator elements $\eta^{i}$ this current only depends on the first prolongation.
\begin{align}
    J_{(\text{Mei}),(1)} =  \textbf{Y}^{[1]} \mathcal{L}
\end{align}

The remaining seven Mei currents are generated by the nontrivial components of the generators and yield non-vanishing conserved quantities:
\bea
    J_{(\text{Mei}),(2)} &=&  \frac{1}{2}r \cdot \textbf{Y}^{[1]} \mathcal{L} + H \frac{2RN^{'2}}{2N} + 2 +6R'N'
\\
    J_{(\text{Mei}),(3)} &=& \Big[N^2 N'(NR' + 2N'R) \Big]^{-1} \nonumber\\
    & & \cdot \Big[8N^4RR^{'2} + 9N^3N'R^2R' + 4H'N^3N'R^{'2} + 12H'N^2N^{'2}RR' \nonumber\\
    & & +4HN^2N^{'2}R^{'2} + 4H'NN^{'3}R^2 + 12H NN^{'3}RR'+HN^{'4}R^2\Big]
\\
    J_{(\text{Mei}),(4)} &=& \frac{NR(8N^2R^{'2} + 5NN^{'}RR'+2N^{'2}R^2)}{N'(NR'+2N'R)}
\\
    J_{(\text{Mei}),(5)} &=& \Big[N^2N^{'2}(4N^2R^{'2}+9NN'RR'+2N^{'2}R^2) \Big]^{-1} \nonumber\\
    & & \cdot \Big[16N^6RR^{'3}+4N^3N^{'3}R^4 +4HNN^{'5}R^3 - 16H'N^5N'R^{'3} \nonumber\\
    & & +3HN^{'6}R^3r -8HN^4N^{'2}R^{'3} + 24N^4N^{'2}R^3R' +36N^5N'R^2R^{'2} \nonumber\\
    & & +34HN^2N^{'4}R^2R' +28H''N^3N^{3'}RR^{'2} + 10H'N^3N^{'3}R^{'3}r \nonumber\\
    & & -12H'N^4N^{'2}RR^{'2}+16HN^3N^{'3}R^{'3}r +16H'N^4N^{'2}R^{'3}r \nonumber\\
    & & +9N^3N^{'3}R^3R^{'3}r +44N^4N^{'2}R^2R^{'2}r + 8H'NN^{'5}R^5r \nonumber\\
    & & +32N^5N'RR^{'3}r +25HN'N^{'5}R^2R'r+56HN^2N^{'4}RR^{'2}r \nonumber\\
    & & +34H'N^2 N^{'4}R^2R'r + 54H'N^3N^{'3}RR^{'2}r\Big]
\\
    J_{(\text{Mei}),(6)} &=& \frac{8N^4R^{'2}-4N^2N^{'2}R^2 +4H'NN^{'3}R - 3HNN^{'3}R' + HN^{'4}R}{NN'(NR'+2N'R)}
\\
    J_{(\text{Mei}),(7)} &=& \frac{2NR(8N^2R^{'2}+5NN'RR'-N^{'2}R^2)}{N'(NR' + 2N'R)}
\\
    J_{(\text{Mei}),(8)} &=& \Big[NN^{'2} (44N^2R^{'2}+9NN'RR'+2N^{'2}R^2) \Big]^{-1} \nonumber\\
    & & \cdot \Big[16N^6R^{'3}+8HNN^{'5}R^2 + 20N^5N'RR^{'2}-HN^{'6}R^{6}r \nonumber\\
    & & +32N^5N'R^{'3}r - 22HN^3N^{'3}R^{'2} + 8H'N^2N^{'4}R^2 -8H'N^4N^{'2}R^{'3}r \nonumber\\
    & & +4N^4N^{'2}R^2R'-4N^2N^{'4}R^3r-16H''N^2N^{'4}R^{'2}r - 2H'N^3N^{'3}R^{'2}r \nonumber\\
    & & -16N^3N^{'3}R^2R'r + 8N^4N^{'2}RR^{'2}r +16HN^2N^{'4}RR' \nonumber\\
    & & +18H'N^3N^{'3}RR'+10H'N^2N^{'4}RR'r - 8HNN^{'5}RR' r\Big]
\eea

\section{Conclusion}
\label{sec:conclusion}

In this work, we performed a comprehensive analysis of symmetry principles in the context of \(f(R)\)-gravity applied to the Reissner--Nordstr\"om (RN) black hole, using a canonical formulation of the Lagrangian. By combining Noether and Mei symmetry approaches, we established a unified framework for deriving symmetry conditions, constructing symmetry generators, and identifying conserved quantities in higher-order gravitational theories.

We first employed \textbf{Noether's theorem} to systematically derive symmetry conditions for the RN metric. Two equivalent derivation schemes were used: the classical variational formulation and the Lie derivative formulation. The latter naturally classifies Noether symmetries into three types:
(i) \textit{general Noether symmetries},
(ii) \textit{canonical Noether symmetries}, and
(iii) \textit{internal Noether symmetries}.
For the quadratic model \(f(R)=R^2\), we obtained explicit symmetry generators such as
\(\mathbf{Y}_1=\partial_r\),
\(\mathbf{Y}_2=r\partial_r + H\partial_H\), and
\(\mathbf{Y}_3=H\partial_H+N\partial_N-R\partial_R\),
which yield conserved currents associated with radial translations, scale transformations, and curvature-metric couplings, respectively.

Beyond Noether symmetries, we applied the \textbf{Mei symmetry approach}, which generalizes the analysis by requiring invariance of the Euler--Lagrange equations rather than invariance of the Lagrangian itself. This framework is particularly powerful for higher-order Lagrangians and systems with non-conservative dynamics. Using the first prolongation of the symmetry generator, we derived a complete set of Mei symmetry conditions for the canonical Lagrangian of the RN metric.

For the quadratic model \(f(R)=R^2\), we obtained eight independent Mei symmetry generators, including \(\mathbf{Y}_1=\partial_r\) and \(\mathbf{Y}_2=\tfrac{1}{2}r\partial_r + H\partial_H\), which recover the scaling and translational symmetries from the Noether sector, but also revealed additional symmetries absent in the standard variational approach. Importantly, we demonstrated that Mei symmetries produce new conserved charges, extending the range of physically meaningful invariants in \(f(R)\)-gravity.

As an application, we computed an energy-like conserved charge associated with the Mei generator \(\mathbf{Y}_1 = 6H\,\partial_H\), obtaining:
\be
    Q_1 = 6H f_R N',
\ee
which establishes a direct correspondence between Mei symmetries and energy conservation in charged black hole spacetimes. This result illustrates how Mei symmetries provide an alternative pathway to conserved quantities in modified gravity.
As summarized in

\subsection*{Future Directions}

This study provides a foundation for symmetry-based analyzes of extended gravitational theories. Promising avenues for future work include the following:

\begin{itemize}
    \item Extending the framework to higher-order curvature models such as \(f(R,\Box R)\) and scalar-tensor extensions \(f(R,\phi,X)\).
    \item Investigating the thermodynamic implications of the Mei-conserved charges for black hole entropy and horizon stability.
    \item Applying the methodology to rotating or dynamical black hole spacetimes, where additional symmetry structures are expected to emerge.
    \item Exploring the relation between Mei symmetries and hidden geometric symmetries, potentially revealing deeper integrability structures in modified gravity.
\end{itemize}

In summary, this work demonstrates that Mei symmetries provide a powerful generalization of Noether's theorem, uncovering new conserved quantities and deepening our understanding of the connection between symmetry, dynamics, and conservation laws in \(f(R)\)-gravity. These findings pave the way for broader applications of symmetry-based techniques in black hole physics and beyond.


\end{document}